\documentclass{article}
\usepackage{spconf,amsmath,graphicx}
\usepackage{booktabs}    
\usepackage{float}
\usepackage{cite}
\usepackage{graphicx}
\usepackage{amssymb}
\def\x{{\mathbf x}}

\title{Deep Space Separable Distillation for Lightweight Acoustic Scene Classification}
\name{ShuQi Ye, Yuan Tian}
\address{South China University of Technology \\
Department of Electronic Business, School of Electronic and Information Engineering \\ Guangzhou, China}
\begin{document}
\maketitle
\begin{abstract}
Acoustic scene classification (ASC) is highly important in the real world. Recently, deep learning-based methods have been widely employed for acoustic scene classification. However, these methods are currently not lightweight enough as well as their performance is not satisfactory. 
To solve these problems, we propose a deep space separable distillation network. Firstly, the network performs high-low frequency decomposition on the log-mel spectrogram, significantly reducing computational complexity while maintaining model performance. Secondly, we specially design three lightweight operators for ASC, including Separable Convolution (SC), Orthonormal Separable Convolution (OSC), and Separable Partial Convolution (SPC). These operators exhibit highly efficient feature extraction capabilities in acoustic scene classification tasks. The experimental results demonstrate that the proposed method achieves a performance gain of 9.8\% compared to the currently popular deep learning methods, while also having smaller parameter count and computational complexity.


\end{abstract}
\section{INTRODUCTION}
\label{sec:intro}

\begin{figure}[htb]

\begin{minipage}[b]{1.0\linewidth}
  \centering
  \centerline{\includegraphics[width=8.5cm]{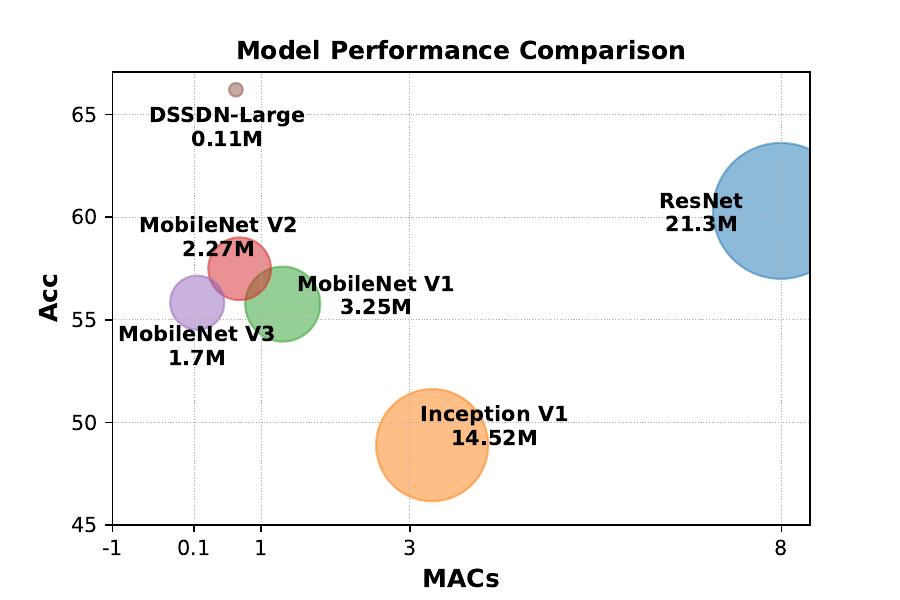}}
\end{minipage}
\caption{\textbf{Model performance comparison. This figure shows the accuracy, computational complexity, and model size of the three lightweight networks we proposed compared with common lightweight networks.}}
\label{figure:0}
\end{figure}
\begin{figure*}[hbt!]
\centering
\includegraphics[width=1\textwidth]{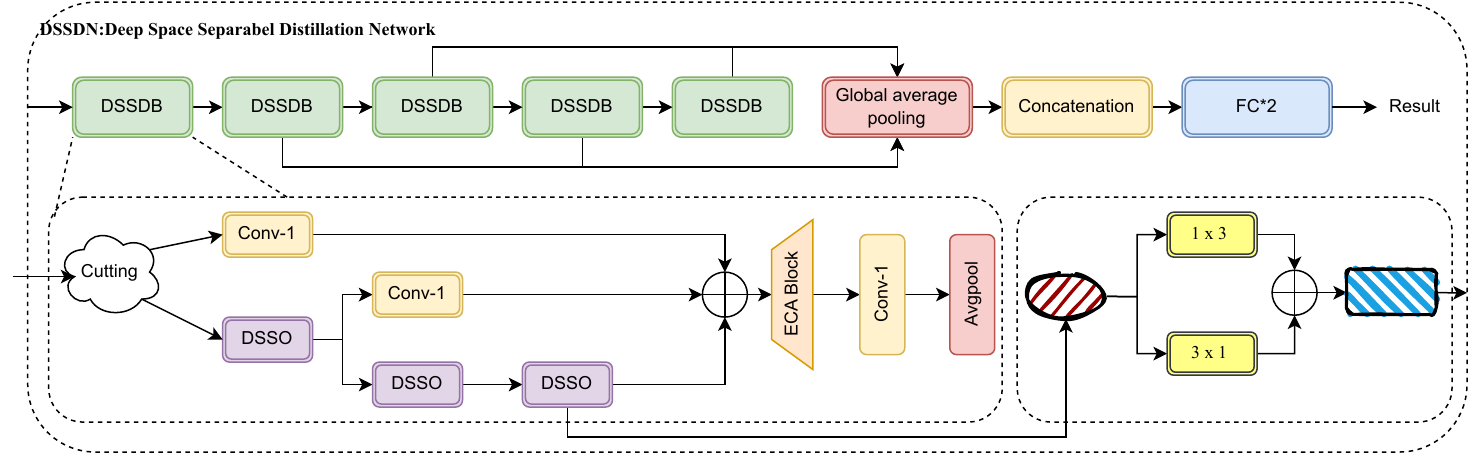}
\caption{\textbf{llustration of the proposed DSSDN framework. The backbone architecture of the proposed DSSDN is composed of five basic modules DSSDB stacked together, and then the channel splicing method is used to fuse the different scale information of the high-level and low-level networks processed by the five modules. DSSDB is built with DSSO as the basic unit, combining the characteristics of the log-Mel spectrum and the characteristics of the distillation block structure, and cutting the frequency axis.}}
\label{figure:1}
\end{figure*}

Acoustic scene classification (ASC) is a method that uses acoustic signal processing technology to identify and classify different acoustic environments. It is widely used in multiple fields such as smart homes, smart transportation, and smart cities. In smart cities, ASC can identify sound scenarios such as commercial areas, residential areas, and parks to achieve automatic city management and improve the quality and efficiency of urban life. In addition, the classification of acoustic scenes is also widely used in the 
fields of voice recognition, voice synthesis, and voice enhancement. Recognition of sound scenes can lead to better understanding of speech signals and improved performance of related tasks. Therefore, the study of the classification of acoustic scenes has extensive application prospects and research value, and it is of great significance to promote the development of artificial intelligence and big data. 

In the field of acoustic scene classification (ASC), research work has achieved some important results. Based on deep learning methods, such as convolutional neural networks (CNN) \cite{DBLP:journals/corr/OSheaN15} and circulating neural network (RNN) \cite{XIE2022103450}, with their strong characteristic learning ability and generalization ability, significant performance improvement has been achieved in the ASC task. Nonetheless, these methods usually require a large amount of training data and computing resources, resulting in too many model parameters, high model complexity, and not lightweight enough. On the other hand, based on traditional signal processing techniques, such as Mel-Frequency Cepstral Coefficient (MFCC) \cite{9955539} and Short-time Fourier transform (STFT), although the calculation complexity is low, the performance is often not as good as deep learning methods. Therefore, in order to achieve high-performance and lightweight balance in the ASC task, we need to seek more lightweight algorithms.

In order to reduce the calculation burden of the model, some lightweight networks have been proposed, such as MobileNet \cite{DBLP:journals/corr/HowardZCKWWAA17}, Shufflenet \cite{DBLP:journals/corr/ZhangZLS17}, and Squeeeznet \cite{DBLP:journals/corr/IandolaMAHDK16}. These networks effectively reduce the computing complexity of the model by simplifying the convolutional layer, reducing the number of channels and reducing resolution. However, the performance of these lightweight networks is often inferior to traditional deep convolutional neural networks (CNN).
On the other hand, technologies such as knowledge distillation \cite{hinton2015distilling} and model compression \cite{DBLP:journals/corr/abs-2105-10059} are also widely used in lightweight networks to improve the performance of the model. Knowledge distillation imitates a larger "teacher" network by training a smaller "student" network to achieve model compression. However, this method often requires a lot of training data and computing resources. Model compression technology, such as weight trimming and weight weight, can directly compress the pre -training model, thereby reducing the size and computing complexity of the model. However, these methods may lead to a decline in model performance.

In summary, although the existing lightweight networks have achieved certain results in reducing the complexity of the model, there are still many shortcomings. These methods are currently not lightweight enough and their performance is not satisfactory in the scene of acoustic scene classification. Therefore, we need to study a more efficient and better lightweight network to meet the needs of acoustic scene classification tasks and improve the performance and efficiency of acoustic scene classification tasks.

We design three lightweight operators specifically for the task characteristics of acoustic scene classification, and build three lightweight networks using the three operators as basic units. The results of the model operation are compared with common lightweight networks such as ResNet. The results are shown in Figure \ref{figure:0}. It can be seen that the three networks we proposed are significantly better in performance.

Our contributions are summarized as follows:

$\bullet$ We propose a new method to cut the frequency axis so that the low-frequency features are continuously strengthened while the high-frequency features can be concentrated, based on the discovery that the characteristics of the log-Mel spectrum on the frequency axis play an important role in the acoustic scene classification task.

$\bullet$ We design a lightweight operator called Orthonormal Separable Convolution to apply regularization to the network architecture of separable convolution. For models with a high number of channels, this operator can effectively reduce the number of parameters and improve model accuracy.

$\bullet$ We also design a lightweight operator called Separable Partial Convolution, which applies partial convolution to the network architecture of separable convolution. It is suitable for models with a small number of channels.

$\bullet$ Compared with traditional lightweight networks, the lightweight operator we design can better meet the task requirements of acoustic scene classification and has better performance.
\section{METHOD}
\label{sec:format}

\subsection{Overview}
\label{ssec:subhead}

The overall structure of our network is shown in Figure \ref{figure:1}. Our network is founded on the blueprint separable residual network structure presented in \cite{9857155} and has been refined. To mitigate the complexity of the model, we introduce a Deep Space Separable Distillation Network (DSSDN). We first construct the fundamental component, the Deep Space Separable Operator (DSSO), and subsequently build a basic module, the Deep Space Separable Distilled Block (DSSDB), based on this fundamental component. We stack these basic modules to form our final model, DSSDN. The network architecture processes the input log-spectrogram through the model's backbone, which consists of five basic module cascade reactions. We then employ a channel splicing technique to integrate different scale information from both high-level and low-level networks. This strategy boosts the performance of the model as it enables the model to better capture features at various scales.


\subsection{Deep Space Separable Distillation Block (DSSDB)}
\label{ssec:subhead}
The overall
structure of DSSDB is shown in the lower left corner of Figure \ref{figure:1}. Based on a tree structure in \cite{Wang2023}, for an input feature map, the above-mentioned modules can
perform distillation and concentration layer by layer, and finally feature enhancement through attention, which can better extract features. On this basis, combined with the characteristics of the log-Mel spectrum and the characteristics of the distillation block structure, we cut the frequency axis. For the low-frequency part, its characteristics are continuously strengthened, after processing by 1 $\times$ 1 convolution kernel. This module can decrease the number of
parameters and complete feature distillation. While for the
high-frequency part, its features can be concentrated. In order to enhance the effect of feature concentration, we propose three new models in parallel. The detailed structure of the three new models will be introduced below. The output of the low-frequency part and the high-frequency part are spliced together to provide richer feature information for subsequent layers.

Wang et al\cite{DBLP:journals/corr/abs-1910-03151} propose ECA attention, which is an efficient one-dimensional attention mechanism. Compared with the commonly used two-dimensional attention, the amount of parameters and calculations used are relatively small.

In general, such a structure reduces the number of parameters
and MACs a lot. Moreover, since the high and low frequencies
are separated, fewer features need to be extracted for each
convolution. Hence, it requires fewer weight bits.

\subsection{Separable Convolution (SC)}
\label{2.3:ssec:subhead}

\begin{figure}[htb]

\begin{minipage}[b]{1.0\linewidth}
  \centering
  \centerline{\includegraphics[width=8.5cm]{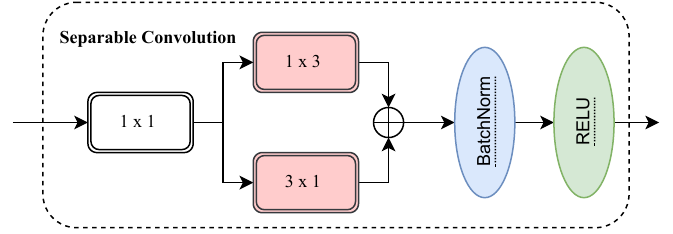}}
\end{minipage}
\caption{\textbf{Structure of Separable Convolution. 
The separable convolution block refers to a model that consists of two distinct channel-wise convolutions, each with kernels of size 3 $\times$1 and 1 $\times$ 3}}
\label{figure:2}
\end{figure}


The separable convolution block \cite{Puy2021} is a newly proposed network model that uses asymmetric convolution, fully combines the advantages of depth and space-separable convolution, and reduces the number of parameters. We migrate the network architecture of separable convolution to our operator, and design the first DSSO, as is shown in Figure \ref{figure:2}, and build the DSSDN-Large architecture based on this operator.

Let $\alpha  \epsilon  R^{k \times c_{in}}$ denotes a feature map of spatial dimension $k$ with $c_{in}$ channels, one layer to the next satisfy:

\begin{equation}
\beta = f(\alpha) = s_{3 \times\ 1} (p _{1 \times\ 1} (\alpha)) + s_{1 \times\ 3} (p _{1 \times\ 1} (\alpha)),
\end{equation}
where $p_{1 \times 1}$ : $R^{k \times c_{in}} \to R^{k \times c_{out}}$ denotes a regular convolution
layer with a kernel of size 1 $\times$ 1 giving a feature map
 $\beta$  of spatial dimension $k$ with $c_{out}$ channels, and $s_{3 \times 1}$, $s_{1 \times 3}$ : $R^{k \times c_{out}} \to R^{k \times c_{out}}$ denote two distinct channel-wise convolutions, each with kernels of size 3 $\times$ 1 and 1 $\times$ 3. $f(\alpha)$ : denote a separable convolution layer, which performs a 1 $1 \times 1$ convolution operation on the input feature map $\alpha$, and then performs convolution operations in two different directions (3 $\times$ 1 and 1 $\times$ 3), and finally adds the outputs of the two directions.

\subsection{Orthonormal Separable Convolution (OSC)}
\label{ssec:subhead}

\begin{figure}[htb]

\begin{minipage}[b]{1.0\linewidth}
  \centering
  \centerline{\includegraphics[width=8.5cm]{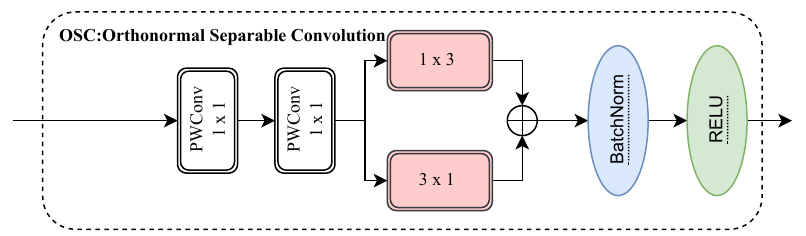}}
\end{minipage}
\caption{\textbf{Structure of OSC. OSC regularizes the 1 $\times$ 1 convolution in the separable convolution.}}
\label{figure:3}
\end{figure}
Daniel \cite{DBLP:journals/corr/abs-2003-13549} proposes that regularizing the conventional convolution can further reduce the parameters. Inspired by previous research, we apply orthogonal regularization to the network architecture of Separable Convolution and proposed the second DSSO, Orthonormal Separable Convolution (OSC) , and build the DSSDN-Small architecture based on this operator to explore whether the number of parameters can be further reduced (see Figure \ref{figure:3} for a visualization) .

One layer to the next satisfy:
\begin{equation}
\beta = s_{3 \times\ 1} (n_{1 \times\ 1} (m_{1 \times 1}(\alpha))) + s_{1 \times\ 3} (n_{1 \times\ 1} (m_{1 \times 1}(\alpha))),
\end{equation}
where $m_{1 \times 1}$ : $R^{k \times c_{in}} \to R^{k \times mid\_channels}$ denotes a convolution layer with a kernel of size 1 $\times$ 1 giving a feature map with $mid\_channels$ channels, $n_{1 \times 1}$ : $R^{k \times mid\_channels} \to R^{k \times c_{out}}$ denotes a convolution layer with a kernel size of 1 $\times$ 1 giving a feature map with $c_{out}$ channels.

\subsection{Separable Partial Convolution (SPC)}
\label{ssec:subhead}

\begin{figure}[htb]

\begin{minipage}[b]{1.0\linewidth}
  \centering
  \centerline{\includegraphics[width=8.5cm]{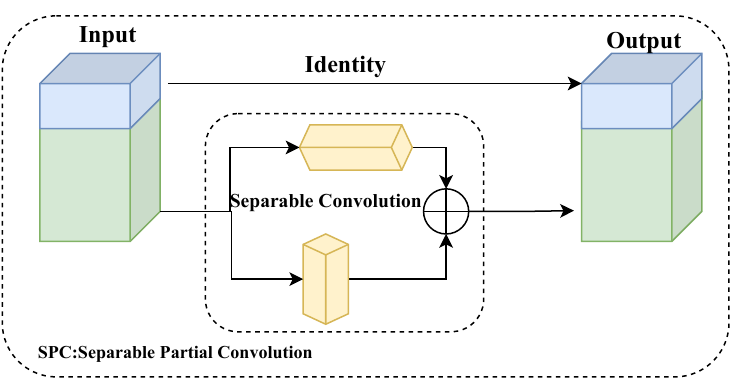}}
\end{minipage}
\caption{\textbf{Structure of SPC. SPC inputs a portion of channels in the feature map into the separable convolution layer for processing.}}
\label{figure:4}
\end{figure}

In order to further optimize the calculation amount and accuracy, we propose another DSSO named Separable Partial Convolution (SPC) and build the DSSDN-Middle architecture based on SPC. As shown in Figure \ref{figure:4}. SPConv's network architecture is improved from PConv\cite{chen2023run} .

PConv simply applies a regular Convolution layer to only a portion of the input channels for spatial feature extraction, leaving the remaining channels untouched. For the channel to be processed, a 3 $\times$ 3 convolutional layer is used. To decrease the number of parameters of the PConv, we replace the regular 3 $\times$ 3 convolutional layer of PConv with Separable Convolution layer in \ref{2.3:ssec:subhead} and name the third DSSO Separable Patrial Convolution. In other words, SPC Combines the advantages of pconv and separable convolution. To fully and efficiently utilize the information from all channels, we further incorporate a pointwise convolution to our SPC. And we build the DSSDN-Middle architecture based on SPC.

The specific calculation formula is as follows:

\begin{equation}
\beta = p_{1 \times 1}(f(x^{in}) \bigcup {\Tilde{x}}^{in}),
\end{equation}
where $\alpha = \x^{in} \bigcup \Tilde{x^{in}}$, $x^{in} \epsilon  R^{k \times c_{dim}}$ denotes a feature map of spatial dimension $k$ with $c_{dim}$ channels, ${\Tilde{x}}^{in} \epsilon  R^{k \times c_{untouched}}$ denotes a feature map of spatial dimension $k$ with $c_{untouched}$ channels, $f(x^{in})$ : $R^{k \times c_{dim}} \to R^{k \times dim}$ denotes a separable convolution layer with kernels of size 3 × 1 and 1 × 3.

\section{EXPERIMENTS}
\label{sec:pagestyle}

\subsection{DATA INTRODUCTION}
\label{ssec:subhead}
To evaluate the performance of DSSC-Large, DSSC-Middle, and DSSC-Small, we conduct a large number of experiments. These experiments involve classifying the sounds collected from different devices into ten predefined categories in order to compare with other methods in terms of parameters, MACs, and accuracy. The dataset used in this paper is described in detail below.

The TAU Urban Acoustic Scenes 2020 Mobile Data Set\cite{heittola2020acoustic} includes audio recordings from 12 European cities captured with 4 different devices. Additionally, the dataset also simulates the data of 11 mobile devices. The development set of the dataset encompasses recordings from 10 cities and 9 devices, of which 3 devices are real and 6 devices are simulated. The evaluation set contains data from 12 cities and 11 devices. There are 2 cities and 6 devices that only exist in the evaluation set, of which 1 device is real and 5 devices are simulated).

\subsection{DATA PREPROCESSING AND AUGMENTATION}
\label{ssec:subhead}

We first pre-process the original audio recording data and convert it to the log-mel spectrogram computed from the com-plete audio signal of 10 seconds, using 2048 points for the FFT, a hop size of 1024, and 256 mel-bins and the size of the generated spectrum is 431 $\times$ 256. And the sampling frequency is set to 44.1 KHz, 
In order to ensure the generalization capacity of the model and prevent overfitting, we adopt two Data augmentation (DA) methods performed in the time frequency: MixuP and SpecAugment. Inspired by \cite{DBLP:journals/corr/abs-2101-04342}, during the training process, we adopt a mix of alpha value of 0.4, and fully train mixed images in the early stage of training and reduce to half in the later stage to reduce the effect of noise sample on the model. At the same time, given that SpacAugment acts on Mel spectrogram rather than audio \cite{park19e_interspeech}, we apply two masking lines on the time domain and frequency domain and the maximum width of each line is 2 to increase the robustness of the network and enhance the specific fragments and specific fragments in the time domain.



In the datasets provided by the competition, most audios are recorded with device A. Therefore the equipment data is unbalanced. We refer to the method of spectrum correction, which is proposed to solve the problem of mismatching record equipment \cite{kosmider20_interspeech}. We first conduct an average of the spectrum of all devices except device A to
obtain a reference device spectrum. Then we use the reference device spectrum to correct the spectrum of device A for additional data. This method is proven to be effective in \cite{kosmider20_interspeech}.

\subsection{TRAINING DETAILS AND EVALUATION INDICATORS}
\label{ssec:subhead}
In this study, we utilize PyTorch as our primary toolkit for conducting all experiments. We employ the stochastic gradient descent (SGD) optimizer and categorical cross-entropy loss function. Our models are trained for 200 epochs with a batch size of 32 and we set the learning rate to 1e-3 and employ the Cosine LRScheduler for updating the learning rate. 

To evaluate the performance of the network architecture comprehensively, we use Accuracy, Parameters and MACs as the evaluation indicators of the experimental results. Among them, the Parameters is used to measure the size of the model, and MACs is used to evaluate the computation of the model. The calculation formula of accuracy is denied as follows:
\begin{equation}
Accuracy=\frac{TP+TN}{TP+TN+FP+FN}.
\end{equation}
where P stands for Positive, N stands for Negative, FP is the count of samples that are in fact negative but predicted as positive, TN is the count of samples that are actually negative and correctly predicted as negative, and TP is the count of samples that are correctly predicted as positive, while FN represents the count of samples that are in fact positive but predicted as negative\cite{9350251}.

\subsection{LIGHTWEIGHT NETWORK COMPARISON EXPERIMENT}
\label{ssec:subhead}
To verify whether the newly proposed network architectures are effective, we compare the three new network architectures with the traditional lightweight network architectures, such as ResNet, MobileNet V1, MobileNet V2, MobileNet V3 and Inception V1.
The recognition results are shown in Table \ref{table:1}. 
Tests on the three new models show that DSSC-Large, DSSC-Middle, and DSSC-Small perform very well. The network parameters of DSSC-Large, DSSC-Middle, and DSSC-Small algorithms are all less than 1M. MACs of them are all less than 0.7G, and the accuracy of them is greater than 65\%. Compared with the five traditional lightweight network models selected in the experiment, the number of parameters and MACs are significantly reduced and the accuracy is significantly improved, thanks to the effectiveness of the deep space separable distillation blocks employed by the three new models. The number of model parameters of DSSC-Large and DSSC-Middle are basically the same, while the number of parameters of DSSC-Small is only 0.078, which is the smallest among all tested models. Compared with DSSC-Large, the number of parameters, MACs and the accuracy of DSSC-Small model are significantly reduced. After adding Orthonormal Regularization to Separable Convolution, it does effectively reduce the number of parameters and MACs, but at the expense of accuracy. The number of parameters, MACs and accuracy of DSSC-Middle are between DSSC-Small and DSSC-Large. The accuracy of the DSSC-Small model is low, and DSSC-Middle uses the SPConv module to increase the number of parameters and MACs in exchange for improved accuracy.

\begin{table}[H]
\caption{\textbf{Experimental results of the comparison experiment}}
\centering
\begin{tabular}{cccc}
\toprule
Models&Parms&MACs&Acc \\
\midrule
ResNet&21.30M&7.99G&60.30 \\
MobileNet V1&3.25M&1.29G&55.76\\
MobileNet V2&2.27M&0.71G&57.48\\
MobileNet V3&1.70M&\textbf{0.14G}&55.83\\
Inception V1&14.52M&3.30G&48.89\\
DSSDN-Large&0.11M&0.66G&\textbf{66.20}\\
DSSDN-Middle&0.11M&0.61G&65.63\\
DSSDN-Small&\textbf{0.08M}&0.56G&65.26 \\
\bottomrule
\end{tabular}
\label{table:1}
\end{table}

\subsection{ABLATION STUDY}
\label{ssec:subhead}
\begin{table}[H]
\caption{\textbf{Experimental results of the ablation study}}
\centering
\begin{tabular}{cccc}
\toprule
Settings&Parms&MACs&Acc \\
\midrule
Baseline&0.11M&0.66G&66.20 \\
DL-O&0.38M&2.45G&55.63 \\
DL-B&0.15M&14.12G&40.90 \\

\bottomrule
\end{tabular}
\label{table:2}
\end{table}

To further validate the effectiveness of our DSSDB module and DSSO module, we utilize the complete DSSDN-Large as the baseline. Subsequently, we substitute the DSSO module and DSSDB module in DSSDN-Large with ordinary 3 $\times$ 3 convolution, respectively. The two new models are generated and named DSSDN-Large without DSSO (DL-O) and DSSDN-Large without DSSDB (DL-B). As shown in Table \ref{table:2}, in comparison to our baseline, the accuracy of DL-O and DL-B decrease by 15.97\% and 38.22\%, respectively. And Parms increase by 245.45\% and 36.36\%, respectively. Macs are respectively rapped up 271.21\% and 2039.4\%. Our findings demonstrate that the DSSDB module effectively distills more features, while the DSSO module contributes to a lighter model.

\section{CONCLUSIONS}
\label{sec:typestyle}
We describe a lightweight system for lightweight acoustic scene classification. We design three lightweight operators as the basic unit of our model. By combining the architecture of DSSDB with three types of DSSO, we propose three deep space separable distillation networks (DSSC-Large, DSSC-Middle, DSSC-Small). Our experiments show that all three new models offer significant improvements in performance. Finally, our three new models achieve accuracy of 66.20\%, 65.63\% and 65.26\%, MACs of 0.66G, 0.61G, 0.56G, and Parms of 0.11M, 0.11M and 0.08M. In other words, the network architecture of DSSDB and DSSO helps to achieve ASC's classification tasks more lighteweight.


 



\bibliographystyle{IEEEbib}
\bibliography{strings,refs}

\end{document}